\documentclass[aps,prc,twocolumn,superscriptaddress,showpacs]{revtex4}
\usepackage{epsfig,pstricks,dcolumn}

\begin{document}

%\preprint{}

\title{Density determinations in   heavy ion collisions}

%% \affiliation command applies to all authors since the last
%% \affiliation command. The \affiliation command should follow the
%% other information
%% \affiliation can be followed by \email, \homepage, \thanks as well.
%\author{J. B. Natowitz}
%%\email{}
%\affiliation{Cyclotron Institute, Texas A \& M University, College Station, Texas 77843-3366, USA}
%\author{R. Wada}
%%\email{}
%\affiliation{Cyclotron Institute, Texas A \& M University, College Station, Texas 77843-3366, USA}
%\author{K. Hagel}
%%\email{}
%\affiliation{Cyclotron Institute, Texas A \& M University, College Station, Texas 77843-3366, USA}
%\author{S. Kowalski}
%%\email{}
%\affiliation{Cyclotron Institute, Texas A \& M University, College Station, Texas 77843-3366, USA}
%\author{L. Qin}
%%\email{}
%\affiliation{Cyclotron Institute, Texas A \& M University, College Station, Texas 77843-3366, USA}

\author{G. R\"{o}pke}
\email{gerd.roepke@uni-rostock.de}
\affiliation{Institut f\"{u}r Physik, Universit\"{a}t Rostock,
Universit\"{a}tsplatz 3, D-18055 Rostock, Germany}
\author{S. Shlomo}
%%\email{}
\affiliation{Cyclotron Institute, Texas A \& M University, College Station, Texas, 77843, USA}
\author{A. Bonasera}
%%\email{}
\affiliation{Cyclotron Institute, Texas A \& M University, College Station, Texas, 77843, USA}
\affiliation{Laboratorio Nazionale del Sud-INFN, v. S. Sofia 64, 95123 Catania, Italy}
\author{J. B. Natowitz}
%%\email{}
\affiliation{Cyclotron Institute, Texas A \& M University, College Station, Texas, 77843, USA}
\author{ S. J. Yennello}
%%\email{}
\affiliation{Cyclotron Institute, Texas A \& M University, College Station, Texas, 77843, USA}
\author{A. B. McIntosh}
\affiliation{Cyclotron Institute, Texas A \& M University, College Station, Texas, 77843, USA}
\author{J. Mabiala}
\affiliation{Cyclotron Institute, Texas A \& M University, College Station, Texas, 77843, USA}
\author{L. Qin}
\affiliation{Cyclotron Institute, Texas A \& M University, College Station, Texas, 77843, USA}
\author{S. Kowalski}
\affiliation{Cyclotron Institute, Texas A \& M University, College Station, Texas, 77843, USA}
\author{K. Hagel}
\affiliation{Cyclotron Institute, Texas A \& M University, College Station, Texas, 77843, USA}
\author{M. Barbui}
\affiliation{Cyclotron Institute, Texas A \& M University, College Station, Texas, 77843, USA}
\author{K. Schmidt}
\affiliation{Cyclotron Institute, Texas A \& M University, College Station, Texas, 77843, USA}
\author{G. Giulani}
\affiliation{Cyclotron Institute, Texas A \& M University, College Station, Texas, 77843, USA}
\author{H. Zheng}
\affiliation{Cyclotron Institute, Texas A \& M University, College Station, Texas, 77843, USA}
\author{S. Wuenschel}
\affiliation{Cyclotron Institute, Texas A \& M University, College Station, Texas, 77843, USA}

\begin{abstract}
The experimental determination of freeze-out   temperatures  and densities from the yields of light elements emitted  in heavy ion collisions is discussed. Results from different experimental approaches are compared with those of model calculations carried out with and without the inclusion of medium effects.  Medium effects  become of relevance for baryon densities  above  $\approx 5 \times 10^{-4}$ fm$^{-3}$. A quantum statistical (QS) model incorporating medium effects is in good agreement with the experimentally derived results at higher densities. A densitometer based on calculated chemical equilibrium constants is proposed. 
%
%Various approaches are presented that determine freeze-out parameter for temperature and density from the yields of light elements observed in heavy ion collisions. A quantum statistical approach is used to calculate cluster yields.
%Medium effects on the equation of state are taken into account that become of relevance for baryon densities $\rho_B \ge  5 \times 10^{-4}$ fm$^{-3}$. According to Natowitz {\it et al.}, the chemical constant $K_\alpha$ can be used to determine the density. Other approaches are discussed. Some examples are given that show that the simple nuclear statistical equilibrium has to be improved.
\end{abstract}

\date{\today}
%\keywords{Nuclear matter equation of state, Symmetry
%energy, Cluster formation, Supernova simulations, Low-density nuclear
%matter}
%21.65.Ef, 05.70.Ce, 25.70.-q, 26.60.Kp, 26.50.+x}

\maketitle

%\section{Introduction: Nuclear Statistical Equilibrium}
Heavy ion collisions (HIC) are often used as a tool to investigate
the properties of excited nuclear matter. Measured
yields of different ejectiles as
well as their their energy spectra and their 
 correlations in momentum space can be used
to infer the properties of the emitting source. Despite
the fact that  a great deal of experimental data has been accumulated from
HIC during the last few  decades, reconstruction of
the properties of the hot expanding nuclear system remains a
  difficult task. Two major problems are the
complications inherent in incorporating   non-equilibrium effects  and in  the treatment
of strong correlations that are already present in
equilibrated  nuclear matter.

An often employed  simple approach to handling these effects  is the freeze-out approximation.
Starting from hot dense matter produced in HIC, this approach
assumes the attainment of  local thermodynamic equilibrium after a short
relaxation time. 
Chemical equilibrium may also be
established in the expanding fireball if the collision rates
in the expanding hot and dense nuclear system are above
a critical value.

While more microscopic approaches employing  transport models
that describe the dynamical evolution of the many particle
system are being pursued, a freeze-out approach provides a  very efficient
means to get a general overview of the  reaction.
Such approaches have been applied   in heavy-ion reactions,
to analyze the equation of state of nuclear matter, see
 \cite{Bondorf:1995ua}, but also recently in high-energy experiments (RHIC,
LHC) to describe the abundances of emitted elementary
particles \cite{Munzinger}. Much information on  the symmetry energy,
on phase instability, etc., has been obtained using this concept.

Within the freeze-out approximation, the abundances
of emitted particles and clusters at freeze-out are determined by the
temperature $T$, the baryon density $n_B$, and
the isospin asymmetry $\delta = (n_n-n_p)/n_B$,  which is related to the
total proton fraction $Y_e=(1- \delta)/2$. In this paper we discuss 
the extraction of densities and temperatures  from the measured
yields of ejectiles  in HIC. We focus  on the information content of neutrons ($n$), protons ($p$), deuterons ($d$), tritons
($t$), $^3$He ($h$), and $^4$He ($\alpha$) particles,  emitted in 
 near  Fermi energy reactions. To extract the relevant information  we optimize
the freeze-out approach by including correlations
and density effects using  systematic, consistent quantum statistical approaches.  

We are considering
only the yields $Y_i$ of these particles, the energy spectra are established by 
long-range interactions and will not be discussed
here. It is possible to extend the approach also to other
situations where not only particles with $A \le 4$ are of relevance.
Whereas the asymmetry is easily obtained from
the proton and neutron number of all emitted fragments,
for the determination of the temperature many efforts
have been made. In particular, double ratios have been
considered. We will not discuss these results here. In
contrast, the determination of the density is a serious
problem that has not been solved in a satisfactory manner
until now. We will give the reason and propose a
solution of this problem.

{\it Experiments and Data Analysis Using the Nuclear Statistical Equilibrium (NSE)}. 
The NIMROD collaboration has recently measured yields of light particles in three
different experiments performed at energies near  the
Fermi energy. Collisions of  $^{64}$Zn  projectiles with $^{92}$Mo and $^{197}$Au target nuclei \cite{Kowalski:2006ju} and the collisions $^{70}$Zn+$^{70}$Zn,
$^{64}$Zn+$^{64}$Zn, and $^{64}$Ni+$^{64}$Ni  were studied at $E/A = 35$ MeV/nucleon  \cite{Kohley}.
Collisions of  $^{40}$Ar + $^{112}$Sn, $^{124}$Sn and $^{64}$Zn +
$^{112}$Sn, $^{124}$Sn  \cite{Natowitz} were studied at 47 MeV/ nucleon .
These experiments have been described in several papers \cite{Kowalski:2006ju,Natowitz,Natowitz1,Natowitz2,Aldo} 
where the details are given. 

Our goal is to derive $T, n_B$, and $\delta$ from the five experimental yields,  $Y_p; Y_d; Y_t; Y_h; Y_\alpha$,  of the light charged $Z=1,2$ species (The neutron yields are not   accurately measured  but, under equilibrium assumptions, can be ascertained from the proton yields combined with $t/h$ ratios as indicated below.)
This problem is easily solved in the low-density limit
where the NSE can be applied, i.e. below $n_B \approx 10^{-4}$ fm$^{-3}$
and at moderate temperatures where medium effects
can be neglected. Using the simple relations for the non-
degenerate  ideal mixture of reacting components
\begin{equation}
n_i=\frac{2 s_i+1}{\Lambda_i^3} {\rm e}^{(E_i+ Z_i \mu_p+N_i \mu_n)/T} , 
\end{equation}
where $\Lambda_i^2= 2 \pi \hbar^2/(m_i T)$ denotes the thermal wave length, $m_i$ the mass, $s_i$ the spin, and $E_i$ the binding energy  of the different components, one can construct expressions
that are almost directly related to the different thermodynamic parameters.

In particular the ratio $Y_h/Y_t$ can be used to determine the asymmetry of the nuclear system.
It can also be used to give an estimate of the neutron yield 
\begin{equation}
Y_n=Y_p \frac{Y_t}{Y_h} f_\delta(T)
\end{equation}
where $f_\delta(T)=\exp[(E_h-E_t)/T] [(m_n m_h)/(m_p m_t)]^{3/2}$
is a correction that accounts for the difference in the binding energies of $^3$H and $^3$He. 
For the sake of simplicity we  use 
in the following  the  approximation $m_A=Am$ with the average baryon mass, $m$.

The temperature can be determined by a double ratio
of yields chosen so that the chemical potentials are compensated
in the NSE. According to Albergo \cite{Albergo}  the temperature can
be obtained from
\begin{equation}
\label{4}
T_{\rm HHe}=\frac{14.3\,\, {\text{MeV}}}{\ln \left[ 1.59 
\frac{Y_\alpha Y_d}{Y_t Y_h}\right]}\,.
\end{equation}

Within the NSE framework, knowledge of the temperature allows the extraction
of the baryon density. In \cite{Kowalski:2006ju}, the yield ratio of $^4$He to $^3$H was used to determine the free proton density according to
\begin{equation}
\label{5}
 n_p = 0.62 \times 10^{36}\, T^{3/2} \exp[-19.8/T] \,\frac{Y_\alpha}{Y_t},
\end{equation}
similarly
\begin{equation}
\label{6}
 n_n = 0.62 \times 10^{36}\, T^{3/2} \exp[-20.6/T] \,\frac{Y_\alpha}{Y_h}.
\end{equation}
Here $T$ is the temperature in MeV,  and $n_i$ has units of nucleons/cm$^3$.The total baryon density follows as $n_B=(n_p/Y_p) \sum_i A_i Y_i$.

{\it Consistency Test for the NSE.}
 Note that only ratios of yields of bound states were used to infer the temperature, Eq. (\ref{4}) and the chemical potentials, Eqs. (\ref{5}), (\ref{6}). To infer the thermodynamic parameter, also other ratios can be considered that contain the free nucleon ($p,n$) yields. If we focus on five measured yields $Y_p, Y_d, Y_t, Y_h, Y_\alpha$, we have four ratios that are of relevance to infer the three parameters $T, n_B, \delta$ that characterize the thermodynamic state of the nuclear system. There is one additional degree of freedom that can be used for a consistency check. In particular, we can consider the
ratio
\begin{equation}
\label{7}
R_{\text {test}}=4^\epsilon \left(\frac{27}{16}\right)^{3 \epsilon/2} \frac{3}{4} \left(\frac{8}{9}\right)^{3/2} \frac{Y_\alpha^{2 \epsilon-1} Y_p^\epsilon}{Y_h^{2 \epsilon-1} Y_t^{\epsilon-1} Y_d}
\end{equation}
with
$\epsilon=(E_\alpha+E_d-E_t-E_h)/(2 E_\alpha-E_t-2 E_h)=0.43833$, the prefactor of the yield fraction has the value 1.62796.

From NSE follows $R_{\text {test}}^{\text{NSE}}=1$. This quantity is also easily
determined from  measured yields. In particular, the data obtained in the experiment Ref. \cite{Kowalski:2006ju} give in total (summed over $v_s$) $R_{\text {test}}=1.22$, the data of \cite{Natowitz}  lead to $R_{\text {test}}=1.36$, and the data of \cite{Aldo} to 1.147 (summed over all excitation energies).
Apparently, in comparison with the yields of bound nucleons, the yield $Y_p$ is higher than expected within NSE. 

Different reasons can be given for this deviation:\\  i) The assumption of thermodynamic equilibrium is not realized. One has to investigate the dynamical non-equilibrium expansion of the fireball produced in HIC.\\ ii) The source is more complex.\\ iii) The assumption of an ideal
mixture of colliding, but otherwise non-interacting components
(free nucleons and clusters) must be improved.

We will not discuss how the freeze-out concept has to
be modified when non-equilibrium and finite size effects
are taken into account. 
Rather here we focus on the last point - improving
the approximation of an ideal mixture by considering
effects of correlations in the medium. This can be done within a systematic
quantum statistical approach

\begin{figure}
\begin{center}
 \includegraphics[width=0.4\textwidth,angle=-0]{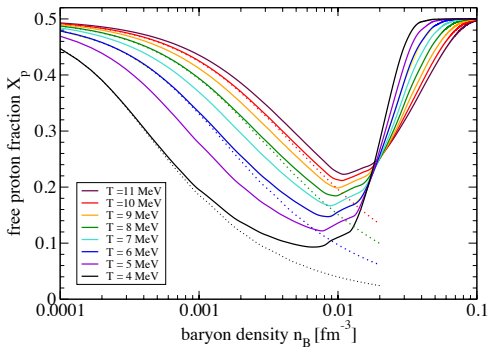}
\end{center}
\caption{\label{Fig.1}%
(Color online) 
Free proton fraction as function of density
and temperature in symmetric matter. Restricting components  to light
elements $A \leq 4$, the QS calculations (solid lines) are compared
with the NSE results (dotted lines). No continuum contributions are included.
The Mott effect  and  its temperature dependence is clearly seen near 0.01 fm$^{-3}$ where the bound state fraction
disappears and the free proton fraction rises.}
\end{figure}

{\it Quantum statistical (QS) approach}. 
Within a quantum statistical approach to nuclear matter, correlations and bound state formation are treated using Green's functions to derive in-medium few-body wave equations, see \cite{Ropke2011}. Comparing to  the zeroth order NSE, improvements are obtained, in particular: \\
i) The classical Boltzmann distribution is replaced by the Fermi or Bose distributions if degenerate effects are to be accounted for. This follows immediately from a quantum statistical approach. In a similar spirit, the momentum quadrupole and normalized number fluctuations for light particle emission in HIC
have been analyzed in Ref. \cite{Aldo}.  In that work it has been proposed
to use the the reduction of fluctuations for Fermi systems or enhancement of fluctuations for Bose systems to estimate the thermodynamic parameters.
\\
ii) With increasing density, medium effects have to be included. Within a quasiparticle picture, the binding energies of the bound states are decreasing with increasing density due to Pauli blocking. Depending on temperature and center of mass momentum, the bound states merge in the continuum at the so-called Mott density. Since the composition is determined by the quasiparticle energies, the cluster abundances are suppressed. As a consequence, the mass fraction of free nucleons is enhanced compared with the NSE, see Fig. \ref{Fig.1}. The medium effects become of relevance when the baryon density $n_B$ exceeds a value of about $5 \times10^{-4}$ fm$^{-3}$. The expressions (\ref{4})~-~(\ref{6}) used to derive the thermodynamic parameters based
on the NSE have to be correspondingly corrected, 
as will be shown in the following. (See also Ref.  \cite{Shlomo:2009}.)\\
iii) In a QS approach, contributions of the continuum
to the density also arise (scattering states). Within a
virial expansion, for each channel where a bound state
is formed, scattering states will also contribute to the 
equation of state. An upper limit for the contributions of
the continuum can be given subtracting for each bound
state the same term with zero binding energy. These
continuum contributions are small in the region considered
here and are neglected in the present work. Future
investigations are needed to account   the continuum
correlations. 

Because the ratio of free nucleons to bound clusters is
strongly influenced by medium effects the use of the NSE is limited to very small densities.
Compared to the NSE, in the QS approach the concentration
of bound states is going down, whereas the fraction of free
nucleons increases. This modifies
the yield fractions that contain the free nucleon yields.

{\it Temperature Determinations in Low Density Nuclear Matter.}
At densities below the Mott point the effect of
medium modifications on the double isotope ratios  is not strong \cite{Shlomo:2009}. Thus, to  
a good approximation the determination of the temperature can be performed employing the double ratios, Eq. (\ref{4}). In Refs. \cite{Kowalski:2006ju,Natowitz,Natowitz1,Natowitz2}  this technique is employed to characterize the temperature evolution of the expanding nascent fireball (the intermediate velocity or nucleon-nucleon source) by associating particle velocity with emission time. (The Albergo expression, Eq. (\ref{4}),  is modified by a factor (9/8)$^{1/2}$ in front
of the double ratio  when applied  to particles with the
same surface velocity, see \cite{Kowalski:2006ju}.) 
In Ref. \cite{Aldo} which focuses on quasiprojectile sources of different excitation energy, temperatures have been calculated
employing  the momentum
quadrupole fluctuation method. In the comparisons which follow, the temperatures are those derived in the quoted references. 

{\it Density Determinations}.
The main problem is the determination of the density
because the influence of medium effects can be  strong.   In the following we compare results from four different approaches to determination of the density,
i) the Albergo NSE based relations \cite{Kowalski:2006ju}, ii) the Mekjian coalescence model which takes into account three body terms which might mimic either a higher density (three body collisions) or Pauli blocking. \cite{Natowitz,Mekjian}, iii) the quantum fluctuation analysis method \cite{Aldo}, and iv) an approach based on use of the  Chemical equilibrium  constant employed in Refs. \cite{Natowitz,Natowitz1,Natowitz2}.
The use of the first three of these techniques to extract temperatures and densities have been well described in the references cited.  
The use of the chemical equilibrium constant, introduced in  \cite{Natowitz}, to characterize the relative yields 
\begin{equation}
 K_c(A,Z) = \frac{\rho(A,Z)}{n_p^Z n_n^{(A-Z)}}\,,
\end{equation}
has some particular advantages.  In
contrast to the free proton fraction, these chemical equilibrium
constants, while  sensitive to the effects of the medium,  are not dependent on the asymmetry
parameter or the choice of competing species present in a model  in the low-density limit where the NSE can be applied. 
Specifically, to infer the values for the thermodynamic
parameters of nuclear matter in HIC at freeze-out from experimental data we  
define the quantity ${ \tilde K}_\alpha$ that is related to the chemical equilibrium 
constant  for $\alpha$  particle formation and can be directly determined from the observed experimental yields,
\begin{equation}
{ \tilde K}_\alpha = \frac{Y_\alpha}{Y_p^4}\frac{Y_h^2}{Y_t^2}\left(\sum_i A_i Y_i\right)^3= \frac{n_\alpha}{n_p^4}\frac{n_h^2}{n_t^2}n_B^3\,.
\end{equation}
The second relation
is found by dividing the particle numbers by their common  volume.
This modified chemical constant  $ { \tilde K}_\alpha  $   does not depend
on the volume of the system. Note that the baryon
density equals $n_B=n_p+n_n+2 n_d+3n_t+3n_h+4n_\alpha$,
if the ejectiles are restricted to $A\leq4$. In general, clusters with
higher  $A$ must included if they are formed from the source 
 under consideration.

Within NSE we can show that 
\begin{eqnarray}
&&\ln {\tilde K}^{\text{NSE}}_\alpha=3 \ln n_B+f_\alpha (T)\,.
\end{eqnarray}
Is applicable  for the low-density region, $f_\alpha (T)=(E_\alpha+2E_h-2E_t)/T+(9/2) \ln [2 \pi \hbar^2/(mT)] -\ln 2$. The 
quasiparticle shifts we have previously calculated  for the single nucleons as well as
for the light clusters \cite{Ropke2011}, indicate  that  medium effects are relevant above the density of about
$n_B = 5 \times10^{-4}$ fm$^{-3}$.
In Figure 2 we present  theoretical values of  $ { \tilde K}_\alpha  $ which have been calculated assuming  including QS
corrections (symmetric matter). The decrease of  $ { \tilde K}_\alpha  $ for densities  above $10^{-2}$ fm$^{-3}$ is due to the
Mott effect that bound states disappear because of Pauli
blocking, see \cite{Natowitz1}.

In our calculations we find essentially  no dependence
on the asymmetry parameter as should be expected for the chemical equilibrium expression. In principle this plot constitutes a densitometer which may be employed to 
estimate the density from experimental yields if the temperature has been determined.
However, in general there
are two solutions so that one has to select out the correct
one. For comparison to the theoretical  values presented in Figure 2 we present also in that figure experimental values for $T = 5$ to 11 MeV, derived from the measured data discussed in Refs. \cite{Natowitz,Natowitz1,Natowitz2}. These data reinforce the interpretation that the natural evolution of the systems under investigation in those works encompasses  densities approaching the Mott point as was previously concluded. 
 \begin{figure}
\begin{center}
 \includegraphics[width=0.4\textwidth,angle=-0]{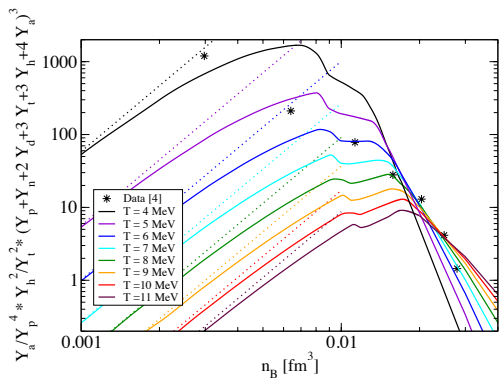}
\end{center}
\caption{\label{Fig.2}%
(Color online) 
Chemical constant $ {\tilde K}_\alpha  $ as function of density and temperature. Data (stars) for $T=5,6,7,8,9,10,11$ MeV \cite{Natowitz} in comparison with the NSE values (thin dotted lines) and QS calculations (bold straight lines).}
\end{figure}

To compare the results of  using  this  densitometer  (number iv) in our list of possible techniques),   with results from the other three techniques in the list we now use comparisons to the theoretical curves to derive densities from the observed experimental values of reference \cite{Natowitz}. These derived values are slightly different than those extracted using a coalescence model. The comparison of   results from   different techniques of extracting $T$ and $n_B$ from   experimental data are presented in Figure 3. 
\begin{figure}
\begin{center}
 \includegraphics[width=0.4\textwidth,angle=-0]{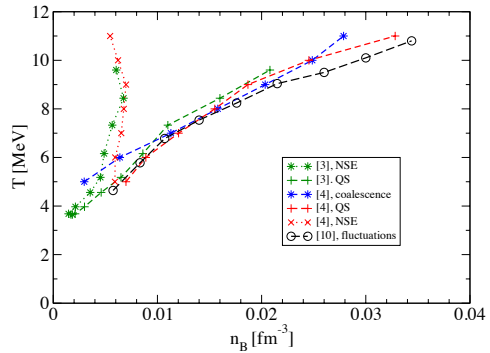}
\end{center}
\caption{\label{Fig.3}%
(Color online) 
Baryon density derived from yields of light elements. Data according to \cite{Kowalski:2006ju,Natowitz,Aldo} are compared with the results of the analysis of yields using NSE and QS calculations for $ {\tilde K}_\alpha  $.}
\end{figure}
%FLIP FIGURE PLOT T on x axis, RHO on y axis.
The use of the basic NSE gives unrealistically low densities reflecting the limitations of that model and its region of applicability \cite{SH,Mekjian,Natowitz} . This point was already apparent in  
the results for laboratory tests of the 
the astrophysical EoS \cite{Natowitz} that also
demonstrate  the relevance of medium effects above  $n_B \approx 10^{-3}$ nuc/fm$^3$.
Interestingly the results of the coalescence model analysis, the quantum fluctuation analysis presented in   Figure 2 lead to very similar results even though different systems and sources have been explored. Both are quite similar to the  densitometer analysis based on QS model results. We return to this point below. 

{\it Discussion.} 
Substantial progress has been made in the effort to explore
nuclear matter at subsaturation densities. There
is now experimental evidence that proves the relevance of
incorporating medium effects such as Pauli blocking and the Mott effect
into theoretical treatments.
As expected from a quantum statistical approach, the
NSE based on non-interacting components is not sufficient 
to explain the data from experiments that investigate
nuclear systems at densities around one tenth of saturation
density and above. Considering the clusters as
quasiparticles, a smooth transition from the NSE at low
densities to mean-field approaches at the saturation density
can be modeled \cite{Typel}. The Albergo densitometer
is  restricted to very low densities. The 
densitometer proposed here, based upon chemical equilibrium constants 
calculated within the framework of the QS model,  can be applied at significantly higher densities.

According to Fig. 2, measured yields of light elements
can be used to infer the baryon density if the temperature
is known. Despite the
double valued solution, this diagram may serve as an important
tool to derive densities from measured yields.
Two other independent methods have been used to infer densities
from the yields of light clusters:\\
i) The Mekjian coalescence model \cite{Mekjian} has been used.  
  Coalescence parameters $P_0$ were calculated for
the different clusters, see \cite{Natowitz2}, and used to determine volumes. The corresponding volume
was used to convert the measured yields into densities.
The results are shown in Figs. \ref{Fig.2} and \ref{Fig.3}. \\
ii) An alternative approach to infer the parameter values
for density and temperature, proposed in
  \cite{Aldo}, employs quadrupole momentum fluctuations and the fluctuations of fermion and boson
numbers in the nuclear system. Compared with classical
systems number fluctuations are decreased for fermion systems
and increased for boson systems if the temperature approaches
the critical temperature. \\
The density values derived by both the coalescence and fluctuation methods   
in rather good agreement   with QS results
that include  medium effects, but in   disagreement with the values derived from
NSE. Only below densities of
about $5 \times 10^{-4}$ fm$^{-3}$, the NSE is applicable.

The  discrepancies
with NSE are substantially reduced  if medium effects such as Pauli blocking \cite{Ropke2011} or, alternatively, excluded
volume \cite{Mekjian,Hempel}  are taken into account.
The fact that the two different experimental results for the temperature and density regions explored are consistent with each other despite the fact that they are obtained from quite different emitting sources  and analyses, suggests that an underlying unifying feature of the EOS is responsible. Indeed, further analysis by Mabiala {\it et al.} \cite{Aldo} indicate that the data are sampling the vapor branch of the liquid gas coexistence curve and within the framework of the Guggenheim systematics may be employed to determine the critical temperatures of mesoscopic nuclear systems,  in a manner analogous to previous treatments \cite{Elliott, N}. 

In conclusion we note some {\it open questions}  with
respect to the present determination of thermodynamic
parameter values:\\
i) The formation of larger clusters is neglected. For their
inclusion see Ref. \cite{Hempel}. The chemical equilibrium constants are not 
sensitive to the formation of other clusters.\\
ii) Continuum correlations are not taken into account.
Possibly they are less important if the quasiparticle picture
and resonances for nearly bound states are included.
A non-equilibrium approach is needed to follow continuum
correlations during the expansion of the fireball.\\
iii) The freeze-out concept is only a simplified approach
and can be improved by more dynamical descriptions of
the non-equilibrium time evolution during HIC.

{\bf Acknowledgement}:
The work was supported by the
US Department of Energy under Grant No. DE-FG03-
93ER40773 and the Robert A. Welch Foundation under
Grant No. A0330.

\end{document}